\documentclass[runningheads,a4paper]{llncs}

\usepackage{times}
\usepackage{hyperref}
\usepackage{xcolor}
\usepackage{amssymb}
\usepackage{stmaryrd}

\usepackage{laws}

\usepackage{imakeidx}
\makeindex
\indexsetup{noclearpage}

\newcommand{\draftonly}[1]{}

\newif\ifArXiV
\ArXiVtrue
\newcommand{\ArXiVAppendix}{\ifArXiV Appendix~\ref{S-proofs}\else \cite{Hayes18fairness-TR}\fi}

\newcommand{\Sect}[1]{Sect.~\ref{#1}}
\newcommand{\Fig}[1]{Fig.~\ref{#1}}

\newcommand{\RvG}{Van Glabbeek}

\newcommand{\Just}{\Keyword{fair}}
\newcommand{\just}{fair}
\newcommand{\JJust}{Fair}
\newcommand{\justparallel}{\mathbin{\Parallel_f}}

\newcommand{\implies}{\mathrel{\Rightarrow}}
\newcommand{\sdefs}{\mathrel{\widehat{=}}}
\newcommand{\refsto}{\mathrel{\sqsubseteq}}
\newcommand{\nat}{\mathbb{N}}
\newcommand{\sync}{\mathbin{\otimes}}

\newcommand{\Identity}[1]{\mathbf{1_{#1}}}
\newcommand{\syncid}{\Identity{}}
\newcommand{\Identitycommand}[1]{\mathit{\mathcal{I}d_{#1}}}
\newcommand{\syncidcommand}{\Identitycommand{}}

\newcommand{\AtomicSteps}{\mathcal{A}}
\newcommand{\Commands}{\mathcal{C}}
\newcommand{\anegate}{\mathop{!}}

\newcommand{\Pset}{\mathop{\mathbb{P}_{}}}

\newcommand{\Keyword}[1]{\mathbf{#1}}
\newcommand{\Do}{\mathop{\Keyword{do}}}
\newcommand{\Od}{{\ \Keyword{od}}}
\newcommand{\Chaos}{\Keyword{chaos}}
\newcommand{\Nil}{\Keyword{nil}}
\newcommand{\Skip}{\Keyword{skip}}
\newcommand{\Term}{\Keyword{term}}
\newcommand{\Finite}{\Fin{\cstepd}}
\newcommand{\Infinite}{\Inf{\cstepd}}
\newcommand{\True}{\mathsf{true}}

\newcommand{\Magic}{\top}
\newcommand{\Abort}{\bot}
\newcommand{\Guard}{\mathrel{\rightarrow}}
\newcommand{\Seq}{\mathbin{\!\mathbin{;}\!}} 
\newcommand{\SSeq}{~}
\newcommand{\Fin}[1]{#1^{\star}}
\newcommand{\Inf}[1]{#1^{\infty}}
\newcommand{\Om}[1]{#1^{\omega}}
\newcommand{\nondet}{\mathbin{\sqcap}}
\newcommand{\Nondet}{\mathop{\bigsqcap}}
\newcommand{\together}{\mathbin{\Cap}}
\newcommand{\Parallel}{\mathop{\parallel}}
\newcommand{\earlyparallel}{\mathbin{\Parallel_e}}

\newcommand{\Pstep}[1]{\Pi(#1)}
\newcommand{\Pstepssp}{\Pstep{\sigma,\sigma'}}
\newcommand{\Estep}[1]{{\cal E}(#1)}
\newcommand{\Estepssp}{\Estep{\sigma,\sigma'}}

\newcommand{\cstepd}{\alpha}
\newcommand{\cpstepd}{\pi}

\newcommand{\cestepd}{\epsilon}

\newcommand{\enabled}[1]{\delta(#1)}

\makeatletter
\newcommand{\ChainRel}[1]{\crcr \noalign{\penalty\interdisplaylinepenalty}
  \hspace*{-1em}#1
  \@ifnextchar*{\@ChainRelCommment}{}}
\newcommand{\ChainRelFormat}[1]{\mbox{\color{blue}~~~#1}}
\def\@ChainRelCommment*[#1]{\ChainRelFormat{#1}
  \crcr \noalign{\penalty\interdisplaylinepenalty}}
\newcommand{\StartRef}[1]{\hspace*{-1.5em}(\ref{#1}) \refsto
  \@ifnextchar[{\@StartRefCommment}{}}
\def\@StartRefCommment[#1]{\mbox{#1}
  \crcr \noalign{\penalty\interdisplaylinepenalty}}
\makeatother

\newcommand{\IFF}{\ChainRel{\Leftrightarrow}}

\newcommand{\ImpliedBy}{\ChainRel{\Leftarrow}}
\newcommand{\Refsto}{\ChainRel{\refsto}}

\newcommand{\Equals}{\ChainRel{=}}

\usepackage[colorinlistoftodos]{todonotes}
\newcommand{\backgroundintensity}{50}

\newcommand\notein[3]
{\todo[inline,linecolor=red,backgroundcolor=#2!\backgroundintensity]
  {#1 says: #3}
}
\newcommand{\ihin}[1]{\notein{IH}{yellow}{#1}}

\newcounter{hours}
\newcounter{minutes}
\newcommand{\printtime}{%
  \ifthenelse{\value{hours}<10}{0}{}\thehours:%
  \ifthenelse{\value{minutes}<10}{0}{}\theminutes}

\begin{document}

\title{Encoding \just ness in a \\ synchronous concurrent program algebra\ifArXiV:\\ extended version with proofs\fi%
\thanks{This work was supported by Australian Research Council (ARC) Discovery Project DP130102901.}
}
\titlerunning{Encoding \just ness in a synchronous concurrent program algebra \draftonly{(\today~\printtime)}}
\author{Ian J. Hayes \and Larissa A. Meinicke}
\authorrunning{Ian J. Hayes \and Larissa A. Meinicke \draftonly{(\today~\printtime)}}
\institute{
The University of Queensland, Brisbane, Queensland, Australia
  \draftonly{\\\vspace*{2ex} \today~\printtime}
}
\maketitle

\vspace{-3ex}
\begin{abstract}
Concurrent program refinement algebra provides a suitable basis for supporting 
mechanised reasoning about shared-memory concurrent programs in a compositional manner,
for example, it supports the rely/guarantee approach of Jones.
The algebra makes use of a synchronous parallel operator 
motivated by Aczel's trace model of concurrency and
with similarities to Milner's SCCS.
This paper looks at defining a form of \just ness 
within the program algebra.
The encoding allows one to reason about the \just\ execution of a single process in isolation
as well as define \just-parallel in terms of a base parallel operator, 
of which no \just ness properties are assumed.
An algebraic theory to support \just ness and \just-parallel is developed.

\end{abstract}

\section{Introduction}

In shared memory concurrency, 
standard approaches  to handling \just ness \cite{Park80,Lehmann1981}
focus on defining a \just\ parallel operator, \mbox{$c \justparallel d$},
that ensures each process gets its \just\ share of processor cycles. 
That complicates reasoning about a single process running as part of a parallel composition
because its progress is determined in part by the \just\ parallel operator.
In this paper we first focus on a single process that is run \just ly with respect to its environment.
That allows one to reason about its progress properties in relative isolation,
although one does need to rely on its environment 
(i.e.\ all processes running in parallel with it)
satisfying assumptions the single process makes about its environment.
\JJust\ parallel composition of processes can then be formulated as (unfair) parallel
composition of fair executions of each of the individual processes
(i.e. $\mbox{\just-execution}(c) \parallel \mbox{\just-execution}(d)$),
where \just-execution of a command is defined below.

\paragraph{Un\just\  parallel.}

For a parallel composition, $c \parallel d$, 
the execution of $c$ may be pre-empted forever by the execution of $d$,
or vice versa.
For example, execution of
\begin{equation}\labelprop{example:term}
  x := 1 \parallel \Do x \neq 1 \Guard y := y + 1 \Od
\end{equation}
with $x$ initially zero 
may not terminate if the right side loop pre-empts the left side assignment forever \cite{Glabbeek2016}.
A minimal \just ness assumption is that 
neither process of a parallel composition can be pre-empted by the other process indefinitely.

\paragraph{Aczel traces.}

The denotational semantics that we use for 
concurrency~\cite{DaSMfaWSLwC} is based on Aczel's
model~\cite{Aczel83,BoerHannemanDeRoever99,DeRoever01}, in which the
possible behaviours of a process, specified by \emph{Aczel traces},
describe both the steps taken by the process itself as well as the
steps taken by its environment. An Aczel trace is a sequence of atomic
steps from one state $\sigma$ to the next $\sigma'$, in which each
atomic step is either a \emph{program step} of the form $\Pstepssp$ or
an \emph{environment step} of the form $\Estepssp$. Parallel
composition has an interleaving interpretation and so program and
environment steps are disjoint. Infinite atomic-step sequences denote
non-terminating executions, and finite sequences are labeled to
differentiate those that (i) \emph{terminate}, (ii) \emph{abort} or
(iii) become \emph{infeasible} after the last atomic step in the
sequence. Abortion represents failure (e.g. failure caused by a
violation of environment assumptions), that may be \emph{refined}
(i.e. implemented) by any subsequent behaviour. Infeasibility may
arise due to conflicting constraints in specifications, and is a
refinement of any subsequent behaviours.
Because each Aczel trace of a process defines both its behaviour as well
as the behaviour of its environment, it is possible to include
assumptions and constraints (including fairness) on the environment of
a process in its denotation -- the set of Aczel traces that it (or any
valid implementation of it) may perform.

When two processes are combined in parallel, each must respect the
environmental constraints placed upon it by the other process --
unless either fails, in which case the parallel composition also fails. 
For example, assuming neither process has failed, one process
may only take a program step $\Pstepssp$ if its parallel process
may perform a step $\Estepssp$,
which permits its environment to take that program step at that point of execution.  
This is achieved by requiring parallel processes to
synchronise on every atomic step they take: 
a program step $\Pstepssp$ of one process matches the
corresponding environment step $\Estepssp$ of the other to give a
program step $\Pstepssp$ of the parallel composition, and identical
environment steps of both processes match to give that environment
step for the parallel composition. Attempting to synchronise other
steps is infeasible.

Let $\cpstepd$ specify the nondeterministic command that executes a
single atomic program step and then terminates, but does not constrain
the state-transition made by that step, that is, $\cpstepd$ could take
$\Pstepssp$ for any possible states $\sigma$ and $\sigma'$.
Similarly, let $\cestepd$ represent the non-deterministic command that
executes any single atomic environment step and then
terminates~\cite{DaSMfaWSLwC,FM2016atomicSteps,FMJournalAtomicSteps-TR}. 
Neither $\cpstepd$ nor $\cestepd$ is allowed to fail: 
they do not contain aborting behaviour.

The command $\Fin{c}$ represents finite iteration of command $c$, zero or more times, and
$\Om{c}$ represents finite or infinite iteration of $c$, zero of more times.
The command $\Inf{c}$ is the infinite iteration of $c$.
Note that $\Om{c}$ splits into finite and infinite iteration of $c$,
where $\nondet$ represents (demonic) nondeterministic choice.
\begin{equation}
  \Om{c} = \Fin{c} \nondet \Inf{c} 
\end{equation}

\paragraph{Imposing \just ness.}

If a process is pre-empted forever its behaviour becomes an
infinite execution of any environment steps, i.e.\ $\Inf{\cestepd}$.
The process $\Just$
that allows any behaviour, except abortion and pre-emption by its environment forever,
can be defined by
\begin{equation}
  \Just \sdefs \Fin{\cestepd} \SSeq \Om{(\cpstepd \SSeq \Fin{\cestepd})}  \labeldef{fair}
\end{equation}
where juxtaposition represents sequential composition.
The process $\Just$ requires all contiguous subsequences of environment steps to be finite.
A process representing \emph{\just\ execution} of a process $c$ is represented by
\begin{displaymath}
  c \together \Just 
\end{displaymath}
where the \emph{weak conjunction}, $c \together d$, of $c$ and $d$ behaves as both $c$ and $d$
unless one of them aborts, in which case $c \together d$ aborts
\cite{AFfGRGRACP,DaSMfaWSLwC}.
Because $\Just$ never aborts, any aborting behaviour of $c \together \Just$ arises solely from $c$.
In this way, $c$ is constrained to be fair until it fails, if ever. 
Weak conjunction is associative, commutative and idempotent;
it has identity $\Chaos$ defined in terms of iteration of any number of atomic steps, 
where $\cstepd$ represents a single atomic step, either program or environment.
\begin{eqnarray} 
  \cstepd & = & \cpstepd \nondet \cestepd \labeldef{alpha} \\
  \Chaos & \sdefs & \Om{\cstepd} \labeldef{chaos} 
\end{eqnarray}
Because program and environment steps are disjoint, the conjunction of these
commands is the infeasible command $\Magic$, i.e.~$\cpstepd \together \cestepd = \Magic$. 

Our interpretation of the execution of the process,
\begin{equation}\labelprop{inc-y-loop}
  \Do \True \Guard y := y+1 \Od~,
\end{equation}
from an initial state in which $y$ is zero allows the loop to be pre-empted forever by its environment
and thus does not guarantee that $y$ is ever set to, say, 7.
In contrast, the \just\ execution of (\refprop{inc-y-loop}),
\begin{equation}
  \Do \True \Guard y := y+1 \Od \together \Just~,
\end{equation}
rules out pre-emption by its environment forever and 
hence ensures that eventually $y$ becomes 7 (or any other natural number).

\paragraph{\JJust\ termination.}

The command $\Term$ allows only a finite number of program steps 
but does not rule out infinite pre-emption by its environment.
It is defined as follows \cite{AFfGRGRACP,DaSMfaWSLwC},
recalling that $\cstepd = \cpstepd \nondet \cestepd$.
\begin{equation}
  \Term \sdefs \Fin{\cstepd} \SSeq \Om{\cestepd} \labeldef{term}
\end{equation}
If $\Term$ is combined with $\Just$, pre-emption by the environment forever is eliminated
giving a stronger termination property that allows only a finite number of both program and environment steps,
see \Lemma{term-fair}.
\begin{eqnarray*}
  \Term \together \Just & = & \Fin{\cstepd}
\end{eqnarray*}
The notation $c \refsto d$ means $c$ is refined (or implemented) by $d$ and is defined by,
\begin{equation}
  c \refsto d \sdefs ((c \nondet d) = c)~. \labeldef{refines-to}
\end{equation}

Hence if $\Term \refsto c$, then $\Term \together \Just \refsto c \together \Just$,
i.e. \just\ execution of $c$ gives strong termination, 
meaning that there are only a finite number of steps overall, both program and environment.

\paragraph{\JJust ness and concurrency.}

Consider the following variation of example (\refprop{example:term}).
\begin{equation}\labelprop{example:fair-term}
  ((x := 1) \together \Just)  \parallel  (\Do x \neq 1 \Guard y := y + 1 \Od \together \Just)
\end{equation}
The \just\ execution of $x := 1$ rules out infinite pre-emption by the right side 
and hence $x$ is eventually set to one,
and hence the right side also terminates
thus ensuring termination of the parallel composition.
Note that
\begin{eqnarray*}
  (c \parallel d) \together \Just \refsto (c \together \Just) \parallel (d \together \Just) 
\end{eqnarray*}
but the reverse refinement does not hold in general
because $(c \parallel d) \together \Just$ does not rule out $c$ being pre-empted forever by $d$
(or vice versa) within the parallel;
it only rules out the whole of the parallel composition from being preempted by its environment forever.

\paragraph{Parallel with synchronised termination.}

The parallel operator $\parallel$ is interpreted as synchronous
parallel for which every step of the parallel (until failure of either
process) must be a synchronisation of steps of its component
processes: a program and environment step synchronise to give a
program step, $\cpstepd \parallel \cestepd = \cpstepd$, two
environment steps synchronise to give an environment step, $\cestepd
\parallel \cestepd = \cestepd$ and both the processes must terminate
together, $\Nil \parallel \Nil = \Nil$.  This is in contrast to the
\emph{early-termination} interpretation of parallel in which, if one
process terminates the parallel composition reduces to the other
process.
The command $\Om{\cestepd}$, referred to as $\Skip$,
\begin{equation}
\Skip \sdefs \Om{\cestepd}   \labeldef{skip}
\end{equation}
is the identity of parallel composition, meaning that it permits any
possible environment behaviour when executed in parallel with any other command, e.g. 
\begin{math}
c \parallel \Skip  =  c ~.
\end{math}
A command $c$ for which
\begin{equation}
c = c\SSeq \Skip \labeldef{healthy-termination}
\end{equation}
is said to be \emph{unconstrained after program termination}. When it
is executed in parallel with another command, then after termination
of $c$, the parallel composition $c \parallel d$ does reduce to the
other command, $d$. If $d$ is also unconstrained after program
termination, then $c\parallel d$ corresponds to the early-termination
interpretation of parallel. Moreover, $c\parallel d$ is then also
unconstrained after program termination, e.g. $c \parallel d =
(c\parallel d) \SSeq \Skip$, see \Lemma{par-skip}.
In this way (\refdef{healthy-termination}) can be perceived as a
healthiness condition, that is preserved by parallel composition of
healthy commands.

The fair execution of any process $c$
constrains the environment, even after the termination of the program
steps in $c$, so that it cannot execute an infinite number of steps in a row,
e.g. $\Term \together \Just = \Fin{\cstepd}$.
This means that it is not healthy (\refdef{healthy-termination}), and so 
for parallel with synchronised termination, 
simply conjoining $\Just$ to both sides of a synchronous parallel can lead to infeasibility.
Consider another of \RvG's examples \cite{Glabbeek2016}:
\begin{equation}\labelprop{example1}
  (x := 1 \together \Just) \parallel (\Do \True \Guard y := y + 1 \Od \together \Just)~.
\end{equation}
The \just\ execution of $x := 1$ rules out infinite pre-emption by the right side loop,
ensuring $x$ is assigned one,
but \just\ execution of $x := 1$ forces termination of the left side,
including environment steps,
which as the right side is non-terminating leads to an infeasible parallel composition.
To remedy this one needs to allow infinite pre-emption of a branch in a \just\ parallel 
\emph{once the command in the branch has terminated}.
For a command $c$ satisfying (\refdef{healthy-termination}) we have that
\begin{equation}
(c \together \Just) \SSeq \Skip
\end{equation}
represents fair execution of $c$ \emph{until program
  termination}. Like the original command $c$, it remains
unconstrained after program termination (i.e. healthy).
For the example above, we have implicitly that $x := 1$ and the loop ($\Do \True \Guard y := y + 1 \Od$)
are unconstrained after program termination,
and so only requiring both branches to execute fairly until program
termination we get
\begin{equation}\labelprop{example2}
  (x := 1 \together \Just) \SSeq \Skip \parallel (\Do \True \Guard y := y + 1 \Od \together \Just)\SSeq \Skip
\end{equation}
which is no longer infeasible, since the second process is allowed to
execute forever after termination of the program steps in the first.

That leads to the following definition for \just\  parallel,
\begin{eqnarray}
c \justparallel d & \sdefs &
  (c \together \Just) \SSeq \Skip \parallel (d \together \Just) \SSeq \Skip \labeldef{fair-parallel}
\end{eqnarray}
which imposes \just ness on $c$ until it terminates, and similarly for $d$.

Our theory of fairness is based on the synchronous concurrent refinement algebra, 
which is summarised in \Sect{S-SCRA} and
\Sect{s:iterations} gives a set of lemmas about iterations in the algebra. 
\Sect{S-fair} gives basic properties of the command $\Just$,
while \Sect{S-fair-concurrency} gives properties of $\Just$ combined with (unfair) concurrency
and \Sect{S-fair-parallel} uses these to derive properties of the \just-parallel operator
which is defined in terms of 
(unfair) parallel (\refdef{fair-parallel}).

\section{Synchronous concurrent refinement algebra}\label{S-SCRA}

The synchronous concurrent refinement algebra is defined
in~\cite{FM2016atomicSteps,FMJournalAtomicSteps-TR}.
In this section we introduce the aspects that are used to define and
reason about \just ness in this paper.
A model for the algebra based on Aczel traces, as discussed in the
introduction, can be found in~\cite{DaSMfaWSLwC}.

A concurrent refinement algebra
with atomic
steps ($\AtomicSteps$), and synchronisation operators parallel
($\parallel$) and weak conjunction ($\together$) is a two-sorted
algebra
\begin{eqnarray*}
(\Commands, \AtomicSteps, \Nondet, \bigsqcup,  ~\Seq\,, \parallel, \together, \anegate\, , \Nil, \cstepd, \Skip, \Chaos, \cestepd)  
\end{eqnarray*}
where the carrier set $\Commands$ is interpreted as the set of
\emph{commands} and forms a complete distributive lattice with meet
($\Nondet$), referred to as \emph{choice}, and join ($\bigsqcup$),
referred to as \emph{conjunction}, and refinement ordering given by (\refdef{refines-to}), 
where we use
$c \nondet d \sdefs \Nondet \{c,d\}$, and $c \sqcup d \sdefs \bigsqcup \{c,d\}$
to represent the meet and join over pairs of elements. The least and
greatest elements in the lattice 
are the aborting command 
$\Abort \sdefs \Nondet{\Commands}$,
and the infeasible command 
$\Magic \sdefs \bigsqcup{\Commands}$,
respectively. The binary operator ``$\,\Seq\,$'', with identity
element $\Nil$, represents \emph{sequential} composition (and satisfies the
axioms listed in \Fig{figure:axioms}), however we abbreviate $c\Seq d$
to $c \SSeq d$ throughout this paper.

For $i\in \nat$, we use $c^{i}$ to represent the fixed-iteration of
the command $c$, $i$ times. It is inductively defined by $c^{0} \sdefs
\Nil$, $c^{i+1} \sdefs c\SSeq c^{i}$.
More generally, fixed-point operators finite iteration ($\Fin{}$),
finite or infinite iteration ($\Om{}$), and infinite iteration
($\Inf{}$) are defined using the least ($\mu$) and greatest ($\nu$)
fixed-point operators of the complete distributive lattice of
commands,
\\[-3ex]
\begin{minipage}[t]{0.5\textwidth}
\begin{eqnarray}
\Fin{c} &\sdefs& (\nu x .\Nil \nondet c \SSeq x) \labeldef{fin-iter-def}\\
\Om{c} &\sdefs& (\mu x . \Nil \nondet c \SSeq x) \labeldef{om-iter-def}
\end{eqnarray}
\end{minipage}%
\begin{minipage}[t]{0.5\textwidth}
\begin{eqnarray}
\Inf{c} &\sdefs& \Om{c} \SSeq \top               \labeldef{inf-iter-def}
\end{eqnarray}
\end{minipage}\\[1ex]
and satisfy the properties outlined in \Sect{s:iterations}.

The second carrier set $\AtomicSteps \subseteq \Commands$ is a
sub-algebra of \emph{atomic step commands}, defined so that
\begin{math}
(\AtomicSteps, \nondet, \sqcup,\, \anegate\, , \top, \, \cstepd)
\end{math}
forms a Boolean algebra with greatest element $\top$ (also the
greatest command), which can be thought of the atomic step that is
disabled from all initial states, the least element $\cstepd$, the
command that can perform any possible atomic step. The negation of an
atomic step $a\in \AtomicSteps$, written $\anegate a$, represents all
of the atomic steps that are not in $a$. Distinguished atomic
step
$\cestepd \in \AtomicSteps$ 
is used to stand for any possible environment step, and its complement,
$\cpstepd \sdefs \anegate \cestepd$,
is then the set of all possible program steps, giving us that $\cstepd
= \cpstepd \nondet \cestepd$.

Both \emph{parallel} composition ($\parallel$) and \emph{weak
  conjunction} ($\together$) are instances of the synchronisation
operator ($\sync$), in which parallel has command identity $\Skip =
\Om{\cestepd}$, and atomic-step identity $\cestepd$; and weak
conjunction has command identity $\Chaos = \Om{\cstepd}$, and
atomic-step identity $\cstepd$.
As well as satisfying the synchronisation axioms from
\Fig{figure:axioms}, a number of additional axioms, also listed in the
figure, are assumed. These include, for example, that both operators
are abort-strict, (\refprop{par-abort}) and (\refprop{conjoin-abort}),
weak conjunction is idempotent (\refprop{conjoin-idempotent}), and
they include assumptions about the synchronisation of atomic steps,
e.g.  (\refprop{par-pi-pi}) and (\refprop{conjoin-pi-env}).

We follow the convention that $c$ and $d$ stands for arbitrary
commands, and $a$ and $b$ for atomic step commands.
Further, subscripted versions of these stand for entities
of the same kind.
We also assume that choice ($\nondet$) has the lowest precedence, and
sequential composition has the highest; and we use parentheses to
disambiguate other cases.

\begin{figure}
\begin{minipage}[t]{0.5\textwidth}
\textbf{Sequential}
\begin{eqnarray}
c_0 \SSeq (c_1 \SSeq c_2) & = & (c_0 \SSeq c_1) \SSeq c_2 \labelprop{seq-assoc} \\
c \SSeq \Nil ~= &c&  =~  \Nil \SSeq c \labelprop{seq-identity} \\
\bot \SSeq c  &=& \bot \labelprop{seq-annihilation-left}
\end{eqnarray}
\end{minipage}%
\begin{minipage}[t]{0.5\textwidth}
\begin{eqnarray}
(\Nondet C) \SSeq d & = & \Nondet_{c \in C} (c \SSeq d)  \labelprop{seq-distr-right}\\
D \neq \emptyset ~\implies~
c \SSeq (\Nondet D) & = & \Nondet_{d \in D}(c \SSeq d) \labelprop{seq-distr-left} 
\end{eqnarray}
\end{minipage} 
\\[+2ex]
\textbf{Synchronisation operators parallel and weak conjunction}
Both parallel ($\parallel$) and weak conjunction ($\together$) are
instances of the synchronisation operator ($\sync$). For parallel we
take the identity command $\syncidcommand$ to be $\Skip$, and atomic-step identity
$\syncid$ to be $\cestepd$, and for weak conjunction we take
$\syncidcommand$ to be $\Chaos$ and $\syncid$ to be $\cstepd$.
\\[-2ex]
\begin{minipage}[t]{0.55\textwidth}
\begin{eqnarray}
c_0 \sync (c_1 \sync c_2) & = & (c_0 \sync c_1) \sync c_2 \labelprop{sync-assoc}\\
c \sync d & = & d \sync c \labelprop{sync-commutative}\\
c \sync \syncidcommand & = & c \labelprop{sync-id}\\
D \neq \emptyset ~\implies~
c \sync (\Nondet D) &=& \Nondet_{d \in D} (c \sync d)
\labelprop{sync-inf-distrib}
\end{eqnarray}
\end{minipage}%
\begin{minipage}[t]{0.45\textwidth}
\begin{eqnarray}
a \sync \syncid & = & a \labelprop{sync-env} \\
\Nil \sync \Nil & = & \Nil \labelprop{sync-nil-nil} \\
\Nil \sync a\SSeq c & = & \Magic \labelprop{sync-nil-atomic} \\
a \sync b &\in & \AtomicSteps \labelprop{par-closure}
\end{eqnarray}
\end{minipage}
\\[-2ex]
\begin{minipage}[t]{1.0\textwidth}
\begin{eqnarray}
(a \SSeq c) \sync (b \SSeq d) & = & (a \sync b) \SSeq (c \sync d)
\labelprop{sync-interchange-seq-atomic} \\
\Inf{a} \sync \Inf{b} & = & \Inf{(a \sync b)} \labelprop{sync-inf}\\
(c_0 \SSeq d_0) \sync (c_1 \SSeq d_1) & \refsto & (c_0 \sync c_1) \SSeq (d_0 \sync d_1)
\labelprop{sync-interchange-seq} 
\end{eqnarray}
\end{minipage}\\[3ex]
\textbf{Additional parallel and weak conjunction axioms} As well as
satisfying the synchronisation axioms the following axioms of parallel
and weak conjunction are assumed to hold.
\\[-2ex]
\begin{minipage}[t]{0.50\textwidth}
\begin{eqnarray}
c \parallel \Abort & = & \Abort \labelprop{par-abort} \\
c \together \Abort & = & \Abort \labelprop{conjoin-abort} \\
c \together c  & = & c \labelprop{conjoin-idempotent}
\end{eqnarray}
\end{minipage}%
\begin{minipage}[t]{0.50\textwidth}
\begin{eqnarray}
\cpstepd \parallel \cpstepd & = & \Magic   \labelprop{par-pi-pi} \\
\cpstepd \together \cestepd & = & \Magic \labelprop{conjoin-pi-env}\\
c \together \cstepd^i & = & c \parallel \cestepd^i   \labelprop{conjoin-par-finite}\\
c \together \Inf{\cstepd} & = & c \parallel \Inf{\cestepd}   \labelprop{conjoin-par-infinite}
\end{eqnarray}
\end{minipage}
\\[-2ex]
\begin{minipage}[t]{1.0\textwidth}
\begin{eqnarray}
(c_0 \together \cstepd^i) \SSeq d_0 \parallel (c_1 \together \cstepd^i) \SSeq d_1 & = & 
((c_0 \together \cstepd^i) \parallel (c_1 \together \cstepd^i)) \SSeq (d_0 \parallel d_1)
\labelprop{sync-initial} \\
(c_0 \together \cstepd^i) \SSeq d_0 \together (c_1 \together \cstepd^i) \SSeq d_1 & = & 
(c_0 \together c_1 \together \cstepd^i) \SSeq (d_0 \together d_1)
\labelprop{conjoin-sync-initial}\\
(c_0 \parallel d_0) \together (c_1 \parallel d_1) & \refsto & (c_0 \together c_1) \parallel (d_0 \together d_1)
\labelprop{conjoin-interchange-par} 
\end{eqnarray}
\end{minipage}

\caption{Axioms for the synchronous concurrent refinement algebra. We
  let $c, d\in \Commands$ be commands, $C,D\in \Pset \Commands $ be
  sets of commands, $a,b\in \AtomicSteps$ be atomic steps, and $i\in
  \nat$ be a natural number.}\label{figure:axioms}
\end{figure}

\section{Properties of iterations}\label{s:iterations}

In this section we outline the iteration properties required in this paper.
Omitted or abbreviated proofs can be found in~\ArXiVAppendix.

First, from \cite{FM2016atomicSteps,FMJournalAtomicSteps-TR}, we have that
the iteration operators satisfy the basic properties listed in
\Fig{figure:iteration}.
\begin{figure}
\begin{minipage}[t]{0.5\textwidth}
\begin{eqnarray}
\Fin{c} &=& \Nil \nondet c \SSeq \Fin{c} \labelprop{finite-unfold}\\
\Om{c} &=& \Nil \nondet c \SSeq \Om{c} \labelprop{omega-unfold}\\
\Om{c} &=& \Fin{c} \nondet \Inf{c} \labelprop{isolation} \\
\Fin{c} &=& \textstyle\Nondet_{i \in \nat} c^i \labelprop{finite-iteration}
\end{eqnarray}
\end{minipage}%
\begin{minipage}[t]{0.5\textwidth}
\begin{eqnarray}
d \nondet c\SSeq x \refsto x &~\Longrightarrow~& \Om{c}\SSeq d \refsto x
\labelprop{omega-induction}\\
x \refsto d \nondet c\SSeq x &~\Longrightarrow~& x \refsto \Fin{c}\SSeq d
\labelprop{finite-induction}\\
c \SSeq \Fin{(d\SSeq c)} &=& \Fin{(c\SSeq d)}\SSeq c \labelprop{finite-leapfrog}\\
c \SSeq \Om{(d \SSeq c)} &=& \Om{(c \SSeq d)} \SSeq c \labelprop{omega-leapfrog}\\
\Om{(c \nondet d)} &=& \Om{c} \SSeq \Om{(d \SSeq \Om{c})} \labelprop{omega-decomposition}
\end{eqnarray}
\end{minipage}%

\caption{Basic properties of iteration operators for commands $c, d, x \in \Commands$.}\label{figure:iteration}
\end{figure}
The following lemma (also from~\cite{FM2016atomicSteps}), captures
that
prefixes of finite iterations of atomic steps
$\Fin{a} \SSeq c$ and $\Fin{b} \SSeq d$ combine in parallel until
either $\Fin{a}$ or $\Fin{b}$ or both complete.  If both $\Fin{a}$ and
$\Fin{b}$ complete together, the remaining commands after the prefixes
run in parallel: $c \parallel d$.  If the first completes before the
second, $c$ runs in parallel with at least one $b$ followed by $d$,
and symmetrically if the second completes before the first.
\begin{lemmax}[finite-finite-prefix]
\[
\Fin{a}\SSeq c \parallel \Fin{b}\SSeq d = 
        \Fin{(a \parallel b)}\SSeq
  ((c \parallel d) \nondet (c \parallel b\SSeq \Fin{b}\SSeq d) \nondet (a\SSeq\Fin{a}\SSeq c \parallel d))
\]
\end{lemmax}

The next lemma is similar to \Lemma*{finite-finite-prefix}, except one
of the prefixes is finite and the other is possibly infinite.
\begin{lemmax}[finite-omega-prefix]
\[
  \Fin{a}\SSeq c \parallel \Om{b}\SSeq d = 
        \Fin{(a \parallel b)}\SSeq
  ((c \parallel d) \nondet (c \parallel b\SSeq \Om{b}\SSeq d) \nondet (a\SSeq\Fin{a}\SSeq c \parallel d))
\]
\end{lemmax}

The following lemma uses the fact that program steps do not
synchronise with other program steps in parallel
(\refprop{par-pi-pi}), to simplify the parallel composition of two
iterations.
\begin{lemmax}[iterate-pi-par-pi]
$\Om{(\cpstepd\SSeq c)} \parallel \Om{(\cpstepd\SSeq d)} = \Nil$
\end{lemmax}

\begin{proof}
The proof uses (\refprop{omega-unfold}), distribution and then
(\refprop{sync-nil-nil}), (\refprop{sync-nil-atomic}) twice, (\refprop{sync-interchange-seq-atomic}),
and (\refprop{par-pi-pi}). 
\qed
\end{proof}

\begin{lemmax}[iterate-pi-sync-atomic]
For either synchronisation operator, $\parallel$ or $\together$, and
atomic step command $a$,
\[
  \Om{(\cpstepd\SSeq c)} \sync a \SSeq d = (\cpstepd \sync a) \SSeq (c \SSeq \Om{(\cpstepd \SSeq c)} \sync d)~.
\]
\end{lemmax}

\begin{proof}
The proof uses (\refprop{omega-unfold}), distribution and then (\refprop{sync-nil-atomic}) and (\refprop{sync-interchange-seq-atomic}).
\qed
\end{proof}

\begin{lemmax}[distribute-infeasible-suffix]
For any synchronisation operator ($\sync$) that is abort strict,
i.e. $(c \sync \bot)=\bot$ for all $c$, then we have that for any
commands $c$, $d$,
\[
  c \sync d\SSeq \Magic = (c \sync d) \SSeq \Magic~.
\]
\end{lemmax}

\begin{lemmax}[infinite-annihilates]
\(
  (c \together \Infinite) \SSeq d_1 = (c \together \Infinite) \SSeq d_2~.
\)
\end{lemmax}

\begin{proof}
The result follows straightforwardly from the fact that weak
conjunction is abort strict (\refprop{conjoin-abort}), $\Infinite =
\Infinite\SSeq \Magic$ from (\refdef{inf-iter-def}) and
\Lemma{distribute-infeasible-suffix}, together with the fact that
$\Magic\SSeq d_1 = \Magic = \Magic \SSeq d_2$ from
(\refprop{seq-distr-right}) by taking $C$ in (\refprop{seq-distr-right}) to be empty.
\qed
\end{proof}
Taking $d_2$ to be $\Nil$ in the above lemma gives 
$(c \together \Infinite) \SSeq d = c \together \Infinite$,
for any $d$.

\begin{lemmax}[sync-termination]
For commands $c$ and $d$ such that $c=c\together \Finite$ and $d = d\together \Finite$,
\begin{eqnarray*}
(c \SSeq \Fin{a} \parallel d \SSeq \Fin{b}) \SSeq (\Om{a} \parallel \Om{b})
& = &
c \SSeq \Om{a} \parallel d \SSeq \Om{b}
\end{eqnarray*}
\end{lemmax}

The following lemma gives us that parallel composition preserves the
healthiness property (\refdef{healthy-termination}).

\pagebreak[3]

\begin{lemmax}[par-skip]
$(c \SSeq \Skip \parallel d \SSeq \Skip) \SSeq \Skip = c \SSeq \Skip \parallel d \SSeq \Skip$
\end{lemmax}

\begin{proof}
Refinement from left to right is straightforward because $\Skip \refsto \Nil$:
\[
  (c \SSeq \Skip \parallel d \SSeq \Skip) \SSeq \Skip
  \refsto
  (c \SSeq \Skip \parallel d \SSeq \Skip) \SSeq \Nil
  =
  c \SSeq \Skip \parallel d \SSeq \Skip~.
\]
Refinement from right to left can be shown as follows.
\[
 \begin{array}{l}
  c \SSeq \Skip \parallel d \SSeq \Skip
 \Equals*[as $\Skip = \Skip \SSeq \Skip$]
  c \SSeq \Skip \SSeq \Skip \parallel d \SSeq \Skip \SSeq \Skip
 \Refsto*[by sync-interchange-seq (\refprop{sync-interchange-seq})]
  (c \SSeq \Skip \parallel d \SSeq \Skip) \SSeq (\Skip \parallel \Skip)
 \Equals*[$\Skip$ is the identity of parallel composition]
  (c \SSeq \Skip \parallel d \SSeq \Skip) \SSeq \Skip
 \end{array}
\]
\qed
\end{proof}

\section{Properties of \just}\label{S-fair}

This section provides a set of properties of the command $\Just$
culminating with \Theorem{fair-termination},
which allows termination arguments to be decoupled from fairness.
The command $\Chaos$ allows any non-aborting behaviour.
If a command refines $\Chaos$, that command is therefore non-aborting.
The command $\Just$ is non-aborting.
\begin{lemmax}[chaos-fair]
$\Chaos \refsto \Just$
\end{lemmax}

\begin{proof}
The proof uses the definition of $\Chaos$ (\refdef{chaos}),
(\refprop{omega-decomposition}),
the property that $\Om{c} \refsto \Fin{c}$, for any command $c$,
and the definition of $\Just$ (\refdef{fair}).
\[
  \Chaos
  =
  \Om{(\cestepd \nondet \cpstepd)}
  =
  \Om{\cestepd} \SSeq \Om{(\cpstepd \SSeq \Om{\cestepd})}
  \refsto
  \Fin{\cestepd} \SSeq \Om{(\cpstepd \SSeq \Fin{\cestepd})}
  =
  \Just
\quad\qed
\]
\end{proof}

\JJust\ execution of a command is always a refinement of the command.
\begin{lemmax}[introduce-fair]
$c \refsto c \together \Just$
\end{lemmax}

\begin{proof}
The lemma holds because $\Chaos$ is the identity of $\together$ and \Lemma{chaos-fair}:
\[
  c = c \together \Chaos \refsto c \together \Just~.
\quad\qed
\]
\end{proof}

\JJust\ execution followed by \just\ execution is equivalent to \just\ execution.
\begin{lemmax}[fair-fair]
$\Just \SSeq \Just = \Just$
\end{lemmax}

\begin{proof}
\noindent
\begin{minipage}[t]{0.44\textwidth}
\[
 \begin{array}{l}
  \Just \SSeq \Just
 \Equals*[by definition of $\Just$ (\refdef{fair})]
  \Fin{\cestepd} \SSeq \Om{(\cpstepd \SSeq \Fin{\cestepd})} \SSeq \Fin{\cestepd} \SSeq \Om{(\cpstepd \SSeq \Fin{\cestepd})}
 \Equals*[by (\refprop{omega-leapfrog})]
  \Om{(\Fin{\cestepd} \SSeq \cpstepd)} \SSeq \Fin{\cestepd}  \SSeq \Fin{\cestepd} \SSeq \Om{(\cpstepd \SSeq \Fin{\cestepd})}
 \Equals*[as $\Fin{c} \SSeq \Fin{c} = \Fin{c}$, for any $c$]
  \Om{(\Fin{\cestepd} \SSeq \cpstepd)} \SSeq \Fin{\cestepd} \SSeq \Om{(\cpstepd \SSeq \Fin{\cestepd})}
 \end{array}
\]
\end{minipage}~~~~~
\begin{minipage}[t]{0.44\textwidth}
\[
 \begin{array}{l}
  \mbox{}
 \Equals*[by (\refprop{omega-leapfrog})]
  \Fin{\cestepd} \SSeq \Om{(\cpstepd \SSeq \Fin{\cestepd})} \SSeq \Om{(\cpstepd \SSeq \Fin{\cestepd})}
 \Equals*[as $\Om{c} \SSeq \Om{c} = \Om{c}$, for any $c$]
  \Fin{\cestepd} \SSeq \Om{(\cpstepd \SSeq \Fin{\cestepd})}
 \Equals*[by definition of $\Just$ (\refdef{fair})]
  \Just
   \quad\qed
 \end{array}
\]
\end{minipage}
\end{proof}

\JJust\ execution of a sequential composition is implemented by \just\ execution of each command in sequence.
\begin{lemmax}[fair-distrib-seq]
$(c \SSeq d) \together \Just \refsto (c \together \Just) \SSeq (d \together \Just)$
\end{lemmax}
\begin{proof}
The proof uses \Lemma{fair-fair} and then interchanges weak conjunction with sequential (\refprop{sync-interchange-seq}).
\[
  (c \SSeq d) \together \Just
 =
  (c \SSeq d) \together (\Just \SSeq \Just)
 \refsto
  (c \together \Just) \SSeq (d \together \Just)
\quad\qed
\]
\end{proof}

The command $\Skip~(=\Om{\cestepd})$ is the identity of parallel composition.
It allows any sequence of environment steps, including $\Inf{\cestepd}$,
but \just\ execution of $\Skip$ excludes $\Inf{\cestepd}$, 
leaving only a finite sequence of environment steps: $\Fin{\cestepd}$.

\pagebreak[3]

\begin{lemmax}[skip-fair]
$\Skip \together \Just = \Fin{\cestepd}$
\end{lemmax}

\begin{proof}
Expanding the definitions of $\Skip$ (\refdef{skip}) and $\Just$ (\refdef{fair}) in the left side to start.
\[
 \begin{array}{l}
  \Om{\cestepd} \together \Fin{\cestepd} \SSeq \Om{(\cpstepd \SSeq \Fin{\cestepd})}
 \Equals*[by \Lemma{finite-omega-prefix}]
  \Fin{\cestepd} \SSeq 
    ((\Nil \together \Om{(\cpstepd \SSeq \Fin{\cestepd})}) \nondet 
     (\Nil \together \cestepd \SSeq \Fin{\cestepd} \SSeq \Om{(\cpstepd \SSeq \Fin{\cestepd})}) \nondet
     (\cestepd \SSeq \Om{\cestepd} \together \Om{(\cpstepd \SSeq \Fin{\cestepd})}))
 \Equals*[by (\refprop{sync-nil-atomic}) and \Lemma{iterate-pi-sync-atomic} and (\refprop{conjoin-pi-env})]
  \Fin{\cestepd} \SSeq 
    (\Nil \nondet \Magic \nondet \Magic)
 \Equals
  \Fin{\cestepd} 
    \quad\qed
 \end{array}
\]
\end{proof}

The command $\Term$ allows only a finite number of program steps 
but does not exclude an infinite sequence of environment steps,
whereas $\Just$ excludes an infinite sequence of environment steps.
When $\Term$ and $\Just$ are conjoined, only a finite number of steps is allowed overall.
\begin{lemmax}[term-fair]
$\Term \together \Just = \Fin{\cstepd}$
\end{lemmax}

\begin{proof}
Note that 
$\Fin{\cstepd} = \Fin{\cstepd} \SSeq \Fin{\cstepd} \refsto \Fin{\cstepd} \SSeq \Fin{\cestepd} \refsto \Fin{\cstepd} \SSeq \Nil = \Fin{\cstepd}$,
and hence $\Fin{\cstepd} = \Fin{\cstepd} \SSeq \Fin{\cestepd}$.
\[
 \begin{array}{l}
   \Term \together \Just = \Fin{\cstepd}
  \IFF*[by the definition of $\Term$ (\refdef{term}) and $\Fin{\cstepd} = \Fin{\cstepd} \SSeq \Fin{\cestepd}$]
   \Fin{\cstepd} \SSeq \Om{\cestepd} \together \Just = \Fin{\cstepd} \SSeq \Fin{\cestepd}
 \end{array}
\]
The fixed point fusion theorem \cite{fixedpointcalculus1995} is applied with 
$F \sdefs \lambda x \cdot x \together \Just$, 
$G \sdefs \lambda x \cdot \Om{\cestepd} \nondet \cstepd \SSeq x$
and
$H \sdefs \lambda x \cdot \Fin{\cestepd} \nondet \cstepd \SSeq x$.
The lemma corresponds to $F(\nu G) = \nu H$, which holds by the fusion theorem if 
$F \circ G = H \circ F$ and $F$ distributes arbitrary nondeterministic choices.
\[
 \begin{array}{l}
  (F \circ G)(x)
 \Equals*[by the definitions of $F$ and $G$]
  (\Om{\cestepd} \nondet \cstepd \SSeq x) \together \Just
 \Equals*[distributing]
  (\Om{\cestepd} \together \Just) \nondet (\cstepd \SSeq x \together \Just)
 \Equals*[by \Lemma{skip-fair} and expanding the definition of $\Just$ (\refdef{fair})]
  \Fin{\cestepd} \nondet (\cstepd \SSeq x \together \Fin{\cestepd} \SSeq \Om{(\cpstepd \SSeq \Fin{\cestepd})})
 \Equals*[by unfolding (\refprop{finite-unfold}) on $\Fin{\cestepd}$ and distribute]
  \Fin{\cestepd} \nondet 
    (\cstepd \SSeq x \together \Om{(\cpstepd \SSeq \Fin{\cestepd})}) \nondet 
    (\cstepd \SSeq x \together \cestepd \SSeq \Fin{\cestepd} \SSeq \Om{(\cpstepd \SSeq \Fin{\cestepd})})
 \Equals*[by \Lemma{iterate-pi-sync-atomic} and $\cstepd \together \cpstepd = \cpstepd$ and (\refprop{sync-interchange-seq-atomic})]
  \Fin{\cestepd} \nondet 
    \cpstepd \SSeq (x \together \Fin{\cestepd} \SSeq \Om{(\cpstepd \SSeq \Fin{\cestepd})}) \nondet 
    \cestepd \SSeq (x \together \Fin{\cestepd} \SSeq \Om{(\cpstepd \SSeq \Fin{\cestepd})})
 \Equals*[distribute and use definition of $\Just$ (\refdef{fair})]
  \Fin{\cestepd} \nondet \cstepd \SSeq (x \together \Just)
 \Equals*[by the definitions of $H$ and $F$]
  (H \circ F)(x)
 \end{array}
\]
Finally $F$ distributes arbitrary nondeterministic choices because for nonempty $C$,
\[
  F(\Nondet C) = (\Nondet C) \together \Just = \Nondet_{c \in C} (c \together \Just) = \Nondet_{c \in C} F(c)~,
\]
and for $C$ empty,
\(
  F(\Nondet \emptyset) = \Magic \together \Just = \Magic = \Nondet_{c \in \emptyset} (c \together \Just) = \Nondet_{c \in \emptyset} F(c)
\)
because $\Chaos \refsto \Just$.
\qed
\end{proof}

We do not build fairness into our definitions of standard sequential programming constructs
such as assignment, conditionals and loops \cite{DaSMfaWSLwC}, 
rather their definitions allow preemption by their environment forever.
Hence any executable sequential program code may be preempted forever.
The command $\Term$ allows only a finite number of program steps 
but also allows preemption by the environment forever.
If a command $c$ refines $\Term$ it will terminate in a finite number of steps 
provided it is not preempted by its environment forever,
and hence \just\ execution of $c$ only allows a finite number of steps
because preemption by the environment forever is precluded by \just\ execution.
That allows one to show termination by showing the simpler property, $\Term \refsto c$, 
which does not need to consider \just ness. 
Existing methods for proving termination can then be used in the context of fair parallel. 
\begin{theoremx}[fair-termination]
If $\Term \refsto c$, then $\Fin{\cstepd} \refsto c \together \Just$.
\end{theoremx}

\begin{proof}
If $\Term \refsto c$, by \Lemma{term-fair} $\Fin{\cstepd} = \Term \together \Just \refsto c \together \Just$.
\qed
\end{proof}

\section{Properties of \just\ and concurrency}\label{S-fair-concurrency}

This section provides a set of properties for combining $\Just$ with (unfair) concurrency,
in particular it provides lemmas for distributing fairness over a parallel composition.
Details of abbreviated proofs can be found in~\ArXiVAppendix.
The following is a helper lemma for \Lemma{fair-par-fair}.

\pagebreak[3]

\begin{lemmax}[fair-par-fair-expand]
$\Just \parallel \Just ~=~ \Fin{\cestepd} \SSeq (\Nil \nondet \cpstepd \SSeq (\Just \parallel \Just))$
\end{lemmax}

\begin{proof}
The proof begins by expanding the definition of $\Just$ (\refdef{fair}),
then uses \Lemma{finite-finite-prefix} and (\refprop{sync-env}),
then \Lemma{iterate-pi-par-pi}, \Lemma{iterate-pi-sync-atomic} and (\refprop{sync-env})
and finally the definition of $\Just$ once more.
\qed
\end{proof}

\JJust\ execution is implemented by \just\ execution of two parallel processes.
\begin{lemmax}[fair-par-fair]
$\Just \refsto \Just \parallel \Just$
\end{lemmax}

\begin{proof}
\[
 \begin{array}{l}
  \Just \refsto \Just \parallel \Just
 \IFF*[by the definition of $\Just$ (\refdef{fair}) and (\refprop{omega-leapfrog})]
  \Om{(\Fin{\cestepd} \SSeq \cpstepd)} \SSeq \Fin{\cestepd} \refsto \Just \parallel \Just
 \ImpliedBy*[by (\refprop{omega-induction})]
  \Fin{\cestepd} \nondet \Fin{\cestepd} \SSeq \cpstepd \SSeq (\Just \parallel \Just) \refsto \Just \parallel \Just
 \end{array}
\]
The above follows by \Lemma{fair-par-fair-expand} by distributing.
\qed
\end{proof}

\JJust\ execution of $c \parallel d$ can be implemented by \just\ execution of each of $c$ and $d$
but the reverse does not hold in general.

\pagebreak[3]

\begin{lemmax}[fair-distrib-par-both]
$(c \parallel d) \together \Just \refsto (c \together \Just) \parallel (d \together \Just)$
\end{lemmax}

\begin{proof}
The proof uses \Lemma{fair-par-fair} and 
then interchanges weak conjunction and parallel (\refprop{conjoin-interchange-par}).
\[
  (c \parallel d) \together \Just
  \refsto
  (c \parallel d) \together (\Just \parallel \Just)
  \refsto
  (c \together \Just) \parallel (d \together \Just)
    \quad\qed
\]
\end{proof}

The following is a helper lemma for \Lemma{fair-par-chaos}.
\begin{lemmax}[fair-par-chaos-expand]
$\Just \parallel \Chaos = \Fin{\cestepd} \SSeq (\Nil \nondet \cpstepd \SSeq (\Just \parallel \Chaos))$
\end{lemmax}

\begin{proof}
The proof uses the definitions of $\Just$ (\refdef{fair}) and $\Chaos$ (\refdef{chaos}) and (\refprop{omega-decomposition}),
then \Lemma{finite-omega-prefix} and (\refprop{sync-env}),
then \Lemma{iterate-pi-par-pi} and \Lemma{iterate-pi-sync-atomic} and (\refprop{sync-env}),
and finally  (\refprop{omega-decomposition}) and definitions (\refdef{fair}) and (\refdef{chaos}).
\qed
\end{proof}

\JJust\ execution in parallel with $\Chaos$ gives a \just\ execution because $\Chaos$ never aborts.
\begin{lemmax}[fair-par-chaos]
$\Just \parallel \Chaos = \Just$
\end{lemmax}

\begin{proof}
The refinement from left to right is straightforward as $\Chaos \refsto \Skip$ and $\Skip$ is the identity of parallel:
\(
  \Just \parallel \Chaos \refsto \Just \parallel \Skip = \Just.
\)
The refinement from right to left uses the definition of $\Just$.
\[
 \begin{array}{l}
  \Just \refsto \Just \parallel \Chaos
 \IFF*[by the definition of $\Just$ (\refdef{fair}) and (\refprop{omega-leapfrog})]
  \Om{(\Fin{\cestepd} \SSeq \cpstepd)} \SSeq \Fin{\cestepd} \refsto \Just \parallel \Chaos
 \ImpliedBy*[by (\refprop{omega-induction})]
  \Fin{\cestepd} \nondet \Fin{\cestepd} \SSeq \cpstepd \SSeq (\Just \parallel \Chaos) \refsto \Just \parallel \Chaos
 \end{array}
\]
The above follows by \Lemma{fair-par-chaos-expand} and distributing.
\qed
\end{proof}

\JJust\ execution of one process of a parallel composition eliminates behaviour $\Inf{\cestepd}$
for that process and hence because parallel compositions synchronise on $\cestepd$ (\refprop{sync-env}),
that eliminates behaviour $\Inf{\cestepd}$ from the parallel composition as a whole,
provided the parallel process does not abort.
Aborting behaviour of one process of a parallel aborts the whole parallel (\refprop{par-abort})
and aborting behaviour allows any behaviour, including $\Inf{\cestepd}$.
\JJust\ execution of $c \parallel d$ can be implemented by \just\ execution of $c$ (or by symmetry $d$).
\begin{lemmax}[fair-distrib-par-one]
$(c \parallel d) \together \Just \refsto (c \together \Just) \parallel d$
\end{lemmax}

\begin{proof}
The proof uses \Lemma{fair-par-chaos},
then interchanges weak conjunction and parallel (\refprop{conjoin-interchange-par})
and finally uses the fact that $\Chaos$ is the identity of weak conjunction.
\[
  (c \parallel d) \together (\Just \parallel \Chaos)
  \refsto
  (c \together \Just) \parallel (d \together \Chaos)
 =
  (c \together \Just) \parallel d
    \quad\qed
\]
\end{proof}

\section{Properties of \just\ parallel}\label{S-fair-parallel}

This section examines the properties of the \just-parallel operator
(\refdef{fair-parallel}), such as commutativity, distribution over
nondeterministic choice and associativity. The first three results
derive readily from the equivalent properties for parallel.

\begin{theoremx}[fair-parallel-commutes]
$c \justparallel d = d \justparallel c$
\end{theoremx}

\begin{proof}
The proof is straightforward from definition (\refdef{fair-parallel}) of \just-parallel 
because (unfair) parallel is commutative. 
\qed
\end{proof}

\begin{theoremx}[fair-parallel-distrib]
$D \neq \emptyset \implies c \justparallel (\Nondet D) = \Nondet_{d \in D} (c \justparallel d)$
\end{theoremx}

\begin{proof}
Let $D$ be non-empty.
\[
 \begin{array}{l}
  c \justparallel (\Nondet D)
 \Equals*[by the definition of $\justparallel$ (\refdef{fair-parallel})]
  (c \together \Just) \SSeq \Skip \parallel ((\Nondet D) \together \Just) \SSeq \Skip
 \Equals*[as non-empty choice distributes over $\together$, sequential composition and parallel]
  \Nondet_{d \in D} (c \together \Just) \SSeq \Skip \parallel (d \together \Just) \SSeq \Skip
 \Equals*[by the definition of $\justparallel$ (\refdef{fair-parallel})]
  \Nondet_{d \in D} (c \justparallel d)
 \quad\qed
 \end{array}
\]
\end{proof}

\begin{theoremx}[fair-par-monotonic]
If $d_1 \refsto d_2$, then $c \justparallel d_1 \refsto c \justparallel d_2$.
\end{theoremx}

\begin{proof}
The refinement $d_1 \refsto d_2$ holds if and only if $d_1 \nondet d_2 = d_1$ and hence,
by \Theorem{fair-parallel-distrib},
\[
 \begin{array}{l}
  c \justparallel d_1 \refsto c \justparallel d_2
 \IFF
  c \justparallel d_1 \nondet c \justparallel d_2 = c \justparallel d_1
 \IFF
  c \justparallel (d_1 \nondet d_2) = c \justparallel d_1
 \end{array}
\]
because $d_1 \nondet d_2 = d_1$ follows from the assumption.
\qed
\end{proof}

\pagebreak[3]

\JJust-parallel retains \just ness for its component processes with respect to the overall environment
even when one component process terminates.
\begin{theoremx}[fair-parallel-nil]
$c \justparallel \Nil = (c \together \Just) \SSeq \Skip$
\end{theoremx}

\begin{proof}
The proof uses the definition of \just\ parallel (\refdef{fair-parallel}),
the facts that $\Nil \together \Just = \Nil$ and $\Skip$ is the identity of parallel composition.
\[
  (c \together \Just) \SSeq \Skip \parallel (\Nil \together \Just) \SSeq \Skip
  =
  (c \together \Just) \SSeq \Skip \parallel \Skip
  =
  (c \together \Just) \SSeq \Skip
    \quad\qed
\]
\end{proof}

While properties such as commutativity and distributivity are
relatively straightforward to verify, associativity of \just-parallel
is more involved.
A property that is essential to the associativity proof is that
fair-parallel execution of two commands not only ensures that each of
its commands are executed fairly
until program termination, 
but also that the whole parallel composition
is executed fairly
until program termination; 
this is encapsulated in \Theorem{absorb-fair-skip},
but first we give lemmas for the finite and infinite cases.

\pagebreak[3]

\begin{lemmax}[introduce-fair-skip]
$c \justparallel d \refsto ((c \justparallel d) \together \Just) \SSeq \Skip$
\end{lemmax}

\begin{proof}
\[
 \begin{array}{l}
  c \justparallel d
  \Equals*[by \Lemma{par-skip} using the definition of fair parallel (\refdef{fair-parallel})]
  (c \justparallel d) \SSeq \Skip
  \Refsto*[by \Lemma{introduce-fair}]
  ((c \justparallel d) \together \Just) \SSeq \Skip
  \quad\qed
 \end{array}
\]
\end{proof}

\begin{lemmax}[finite-absorb-fair-skip]
\[
 \begin{array}{l}
  (((c \together \Finite) \justparallel (d \together \Finite)) \together \Just) \SSeq \Skip
 = 
  (c \together \Finite) \justparallel (d \together \Finite)
 \end{array}
\]
\end{lemmax}

\begin{proof}
The refinement from right to left follows by \Lemma{introduce-fair-skip}.
The refinement from left to right follows.
\[
 \begin{array}{l}
   (((c \together \Finite) \justparallel (d \together \Finite)) \together \Just) \SSeq \Skip
  \Equals*[by the definition of $\justparallel$ (\refdef{fair-parallel})]
  (((c \together \Finite \together \Just) \SSeq \Skip \parallel 
    (d \together \Finite \together \Just) \SSeq \Skip) \together \Just) \SSeq \Skip
 \Refsto*[by \Lemma{fair-distrib-par-both}]
  ((((c \together \Finite \together \Just) \SSeq \Skip) \together \Just) \parallel 
   (((d \together \Finite \together \Just) \SSeq \Skip) \together \Just)) \SSeq \Skip
 \Refsto*[by \Lemma{fair-distrib-seq} and $\together$ idempotent (\refprop{conjoin-idempotent})]
  ((c \together \Finite \together \Just) \SSeq (\Skip \together \Just) \parallel 
   (d \together \Finite \together \Just) \SSeq (\Skip \together \Just)) \SSeq \Skip
 \Refsto*[as $\Skip \together \Just = \Fin{\cestepd}$ by \Lemma{skip-fair}]
  ((c \together \Finite \together \Just) \SSeq \Fin{\cestepd} \parallel 
   (d \together \Finite \together \Just) \SSeq \Fin{\cestepd}) \SSeq \Skip
 \Refsto*[by \Lemma{sync-termination} as $\Skip \parallel \Skip = \Skip$ and $\Skip = \Om{\cestepd}$]
  (c \together \Finite \together \Just) \SSeq \Skip \parallel (d \together \Finite \together \Just) \SSeq \Skip
 \Equals*[by the definition of $\justparallel$ (\refdef{fair-parallel})]
  (c \together \Finite) \justparallel (d \together \Finite)
 \quad\qed
 \end{array}
\]
\end{proof}

\pagebreak[3]

\begin{lemmax}[infinite-absorb-fair-skip]
\[
 \begin{array}{l}
  (((c \together \Infinite) \justparallel d) \together \Just) \SSeq \Skip
 = 
  (c \together \Infinite) \justparallel d
 \end{array}
\]
\end{lemmax}

\begin{proof}
The refinement from right to left follows by \Lemma{introduce-fair-skip}.
The refinement from left to right follows.
\[
 \begin{array}{l}
   (((c \together \Infinite) \justparallel d) \together \Just) \SSeq \Skip
  \Equals*[by the definition of $\justparallel$ (\refdef{fair-parallel})]
  (((c \together \Infinite \together \Just) \SSeq \Skip \parallel 
    (d \together \Just) \SSeq \Skip) \together \Just) \SSeq \Skip
  \Equals*[by \Lemma{infinite-annihilates}]
  (((c \together \Infinite \together \Just) \parallel 
    (d \together \Just) \SSeq \Skip) \together \Just) \SSeq \Skip
  \Refsto*[by \Lemma{fair-distrib-par-one} and $\together$ is idempotent (\refprop{conjoin-idempotent})]
  ((c \together \Infinite \together \Just) \parallel 
    (d \together \Just) \SSeq \Skip) \SSeq \Skip
  \Equals*[by \Lemma{infinite-annihilates}]
  ((c \together \Infinite \together \Just) \SSeq \Skip \parallel 
    (d \together \Just) \SSeq \Skip) \SSeq \Skip
  \Equals*[by \Lemma{par-skip}]
  (c \together \Infinite \together \Just) \SSeq \Skip \parallel 
    (d \together \Just) \SSeq \Skip
  \Equals*[by the definition of $\justparallel$ (\refdef{fair-parallel})]
  (c \together \Infinite) \justparallel d
 \quad\qed
\end{array}
\]
\end{proof}

\begin{theoremx}[absorb-fair-skip]
$((c \justparallel d) \together \Just) \SSeq \Skip = c \justparallel d$
\end{theoremx}

\begin{proof}
The proof decomposes $c$ and $d$ into their finite and infinite components
based on the observation that the identity of ``$\together$'' is $\Chaos$, 
which equals $\Fin{\cstepd} \nondet \Inf{\cstepd}$.
\[
 \begin{array}{l}
  ((c \justparallel d) \together \Just) \SSeq \Skip
 \Equals*[combine $\Finite \nondet \Infinite$ with both $c$ and $d$ using \Theorem{fair-parallel-distrib}]
  (((c \together \Finite) \justparallel (d \together \Finite)) \together \Just) \SSeq \Skip \nondet 
  (((c \together \Finite) \justparallel (d \together \Infinite)) \together \Just) \SSeq \Skip \nondet {} \\
  (((c \together \Infinite) \justparallel (d \together \Finite)) \together \Just) \SSeq \Skip \nondet 
  (((c \together \Infinite) \justparallel (d \together \Infinite)) \together \Just) \SSeq \Skip 
 \Equals*[by \Lemma{finite-absorb-fair-skip} and \Lemma{infinite-absorb-fair-skip}]
  (c \together \Finite) \justparallel (d \together \Finite) \nondet 
  (c \together \Finite) \justparallel (d \together \Infinite) \nondet {} \\
  (c \together \Infinite) \justparallel (d \together \Finite) \nondet 
  (c \together \Infinite) \justparallel (d \together \Infinite)
 \Equals*[merging choices and $\Finite \nondet \Infinite = \Chaos$, the identity of $\together$]
  c \justparallel d
  \quad\qed
 \end{array}
\]
\end{proof}

With these results we can now verify associativity of fair parallel. 

\begin{theoremx}[fair-parallel-associative]
$(c \justparallel d) \justparallel e = c \justparallel (d \justparallel e)$
\end{theoremx}

\begin{proof}
\[
 \begin{array}{l}
  (c \justparallel d) \justparallel e
  \Equals*[by definition of $\justparallel$ (\refdef{fair-parallel})]
  ((c \justparallel d) \together \Just) \SSeq \Skip \parallel (e \together \Just) \SSeq \Skip
  \Equals*[by \Theorem{absorb-fair-skip}]
  (c \justparallel d) \parallel (e \together \Just) \SSeq \Skip
 \Equals*[by definition of $\justparallel$ (\refdef{fair-parallel})]
  ((c \together \Just) \SSeq \Skip \parallel (d \together \Just) \SSeq \Skip) \parallel (e \together \Just) \SSeq \Skip
  \Equals*[by associativity of parallel]
  (c \together \Just) \SSeq \Skip \parallel ((d \together \Just) \SSeq \Skip \parallel (e \together \Just) \SSeq \Skip)
  \Equals*[by definition of $\justparallel$ (\refdef{fair-parallel})]
  (c \together \Just) \SSeq \Skip \parallel (d \justparallel e)
  \Equals*[by \Theorem{absorb-fair-skip}]
  (c \together \Just) \SSeq \Skip \parallel ((d \justparallel e) \together \Just) \SSeq \Skip
  \Equals*[by definition of $\justparallel$ (\refdef{fair-parallel})]
  c \justparallel (d \justparallel e)
  \quad\qed
 \end{array}
\]
\end{proof}

\section{Conclusions}

Earlier work on fairness \cite{Park80,Lehmann1981} focused on defining fairness 
as part of a fair-parallel operator.
The main contribution of this paper is to separate the concerns of fairness and 
the parallel operator.
That allows us to
(i) reason about the fair execution of a single process in isolation,
for example, via \Theorem{fair-termination};
(ii) start from a basis of the (unfair) parallel operator,
which has simpler algebraic properties;
and
(iii) define the fair-parallel operator in terms of the more basic (unfair) parallel operator
and hence prove properties of the fair-parallel operator in terms of its definition.

The first point is important for devising a compositional approach to reasoning about 
the fairness properties of concurrent systems in terms of the fairness properties 
of their components.
The second point allows us to utilise the synchronous concurrent refinement algebra 
\cite{DaSMfaWSLwC,FM2016atomicSteps,FMJournalAtomicSteps-TR}
(which has similarities to Milner's SCCS \cite{Milner83,CaC})
to encode fairness in an existing theory with no built-in fair-parallel operator.
The third point shows that no expressive power is lost compared to starting
with a fair-parallel operator, in fact, there is a gain in expressiveness as
one can define a parallel composition which imposes fairness on only one of its 
components: $((c \together \Just) \SSeq \Skip) \parallel d$.

Overall, these results indicate that a suitable foundation of handling concurrency and fairness 
can start from a theory in which the parallel operator has no built-in fairness assumptions.
The ability to do this derives from the use of a synchronous parallel operator 
motivated by the rely/guarantee approach of Jones \cite{Jones81d,Jones83a,Jones83b}
and Aczel's trace model for that approach \cite{Aczel83,BoerHannemanDeRoever99,DeRoever01}, 
in which environment steps are made explicit.

\paragraph{Acknowledgements.}

This research was supported
Australian Research Council Discovery Grant DP130102901.
Thanks are due to
Robert Colvin,
Rob \RvG,
Peter H\"{o}fner,
Cliff Jones,
and
Kirsten Winter,
for feedback on ideas presented here.
This research has benefited greatly from feedback members of 
IFIP Working Group 2.3 on Programming Methodology,
in particular, at the meeting in Villebrumier.

\bibliographystyle{plain}
\bibliography{ms}

\begin{thebibliography}{10}

\bibitem{fixedpointcalculus1995}
Chritiene Aarts, Roland Backhouse, Eerke Boiten, Henk Doombos, Netty van
  Gasteren, Rik van Geldrop, Paul Hoogendijk, Ed~Voermans, and Jaap van~der
  Woude.
\newblock Fixed-point calculus.
\newblock {\em Information Processing Letters}, 53:131--136, 1995.
\newblock {Mathematics of Program Construction Group.}

\bibitem{Aczel83}
Peter~H.~G. Aczel.
\newblock On an inference rule for parallel composition, 1983.
\newblock
  \url{http://homepages.cs.ncl.ac.uk/cliff.jones/publications/MSs/PHGA-traces.pdf}
  Private communication to Cliff Jones.

\bibitem{DaSMfaWSLwC}
R.~J. Colvin, I.~J. Hayes, and L.~A. Meinicke.
\newblock Designing a semantic model for a wide-spectrum language with
  concurrency.
\newblock {\em Formal Aspects of Computing}, 29:853–--875, 2016.

\bibitem{BoerHannemanDeRoever99}
F.S. de~Boer, U.~Hannemann, and W.-P. de~Roever.
\newblock Formal justification of the rely-guarantee paradigm for
  shared-variable concurrency: a semantic approach.
\newblock In Jeannette Wing, Jim Woodcock, and Jim Davies, editors, {\em FM'99
  -- Formal Methods}, volume 1709 of {\em Lecture Notes in Computer Science},
  pages 1245--1265. Springer Berlin / Heidelberg, 1999.

\bibitem{DeRoever01}
W.-P. de~Roever.
\newblock {\em Concurrency Verification: Introduction to Compositional and
  Noncompositional Methods}.
\newblock Cambridge University Press, 2001.

\bibitem{AFfGRGRACP}
I.~J. Hayes.
\newblock Generalised rely-guarantee concurrency: An algebraic foundation.
\newblock {\em Formal Aspects of Computing}, 28(6):1057--1078, November 2016.

\bibitem{FM2016atomicSteps}
I.J. Hayes, R.J. Colvin, L.A. Meinicke, K.~Winter, and A.~Velykis.
\newblock An algebra of synchronous atomic steps.
\newblock In J.~Fitzgerald, C.~Heitmeyer, S.~Gnesi, and A.~Philippou, editors,
  {\em FM 2016: Formal Methods: 21st International Symposium, Proceedings},
  volume 9995 of {\em LNCS}, pages 352--369, Cham, November 2016. Springer
  International Publishing.

\bibitem{FMJournalAtomicSteps-TR}
I.J. Hayes, L.A. Meinicke, K.~Winter, and R.J. Colvin.
\newblock A synchronous program algebra: a basis for reasoning about
  shared-memory and event-based concurrency.
\newblock Ext. report at \url{http://arxiv.org/abs/1710.03352}, 2017.

\bibitem{Jones81d}
C.~B. Jones.
\newblock {\em Development Methods for Computer Programs including a Notion of
  Interference}.
\newblock PhD thesis, Oxford University, June 1981.
\newblock Available as: Oxford University Computing Laboratory (now Computer
  Science) Technical Monograph PRG-25.

\bibitem{Jones83a}
C.~B. Jones.
\newblock Specification and design of (parallel) programs.
\newblock In {\em Proceedings of IFIP'83}, pages 321--332. North-Holland, 1983.

\bibitem{Jones83b}
C.~B. Jones.
\newblock Tentative steps toward a development method for interfering programs.
\newblock {\em ACM ToPLaS}, 5(4):596--619, 1983.

\bibitem{Lehmann1981}
D.~Lehmann, A.~Pnueli, and J.~Stavi.
\newblock Impartiality, justice and fairness: The ethics of concurrent
  termination.
\newblock In Shimon Even and Oded Kariv, editors, {\em Automata, Languages and
  Programming: Eighth Colloquium Acre (Akko), Israel July 13--17, 1981}, pages
  264--277, Berlin, Heidelberg, 1981. Springer Berlin Heidelberg.

\bibitem{CaC}
A.J.R.G. Milner.
\newblock {\em Communication and Concurrency}.
\newblock Prentice-Hall, 1989.

\bibitem{Milner83}
R.~Milner.
\newblock Calculi for synchrony and asynchrony.
\newblock {\em Theoretical Computer Science}, 25(3):267--310, 1983.

\bibitem{Park80}
David Park.
\newblock On the semantics of fair parallelism.
\newblock In Dines Bj{\o}rner, editor, {\em Abstract Software Specifications},
  volume~86 of {\em Lecture Notes in Computer Science}, pages 504--526.
  Springer Berlin Heidelberg, 1980.

\bibitem{Glabbeek2016}
Rob~J. van Glabbeek.
\newblock Ensuring livenes properties of distributed systems (a research
  agenda).
\newblock Technical report, NICTA, March 2016.
\newblock Position paper.

\end{thebibliography}

\ifArXiV\else\end{document}\fi
\newpage
\appendix

\section{Proofs of selected lemmas}\label{S-proofs}

In this appendix we provide proofs of lemmas from the body of the
paper that were elided for brevity. Additional supporting lemmas of
the synchronous concurrent program algebra are also included.

\vspace{1ex}
\begin{lemmay}[iterate-pi-par-pi]
$\Om{(\cpstepd\SSeq c)} \parallel \Om{(\cpstepd\SSeq d)} = \Nil$
\end{lemmay}

\begin{proof}
\[
 \begin{array}{l}
  \Om{(\cpstepd\SSeq c)} \parallel \Om{(\cpstepd\SSeq d)} 
\Equals*[by  (\refprop{omega-unfold}) ]
  (\Nil \nondet \cpstepd \SSeq c \SSeq \Om{(\cpstepd\SSeq c)})  
        \parallel (\Nil \nondet \cpstepd \SSeq d \SSeq \Om{(\cpstepd\SSeq d)})
\Equals*[since $\parallel$ distributes over $\nondet$]
  (\Nil \parallel \Nil) \nondet
  (\Nil \parallel \cpstepd \SSeq d \SSeq \Om{(\cpstepd\SSeq d)}) \nondet
  (\cpstepd \SSeq c \SSeq \Om{(\cpstepd\SSeq c)} \parallel \Nil) \nondet
  (\cpstepd \SSeq c \SSeq \Om{(\cpstepd\SSeq c)} \parallel 
   \cpstepd \SSeq d \SSeq \Om{(\cpstepd\SSeq d)})
\Equals*[by (\refprop{sync-nil-nil}), (\refprop{sync-nil-atomic}) twice and (\refprop{sync-interchange-seq-atomic})]
 \Nil ~\nondet~
 (\cpstepd \parallel \cpstepd) \SSeq 
 (c \SSeq \Om{(\cpstepd\SSeq c)} \parallel 
  d \SSeq \Om{(\cpstepd\SSeq d)})
\Equals*[by (\refprop{par-pi-pi}) and $\Magic$ is a left annihilator]
 \Nil ~\nondet~ \Magic
\Equals
 \Nil
    \quad\qed
 \end{array}
\]
\end{proof}

\begin{lemmay}[iterate-pi-sync-atomic]
For atomic step command $a$,
\[
  \Om{(\cpstepd\SSeq c)} \sync a \SSeq d = (\cpstepd \sync a) \SSeq (c \SSeq \Om{(\cpstepd \SSeq c)} \sync d)
\]
\end{lemmay}

\begin{proof}
\[
 \begin{array}{l}
  \Om{(\cpstepd\SSeq c)} \sync a \SSeq d 
\Equals*[by  (\refprop{omega-unfold}) ]
  (\Nil \nondet \cpstepd \SSeq c \SSeq \Om{(\cpstepd\SSeq c)}) \sync a \SSeq d
\Equals*[since $\sync$ distributes over $\nondet$]
  (\Nil \sync a \SSeq d) \nondet
  (\cpstepd \SSeq c \SSeq \Om{(\cpstepd\SSeq c)} \sync a \SSeq d)
\Equals*[by (\refprop{sync-nil-atomic}) and (\refprop{sync-interchange-seq-atomic})]
  (\cpstepd \sync a) \SSeq (c \SSeq \Om{(\cpstepd \SSeq c)} \sync d)
    \quad\qed
 \end{array}
\]
\end{proof}

\begin{lemmay}[distribute-infeasible-suffix]
For any synchronisation operator ($\sync$) that is abort strict,
i.e. $(c \sync \bot)=\bot$ for all $c$, then we have that for any
commands $c$, $d$,
\[
  c \sync d\SSeq \Magic = (c \sync d) \SSeq \Magic~.
\]
\end{lemmay}

\pagebreak[4]

\begin{proof}
Let $c$ and $d$ be any commands. 
First we show
\begin{eqnarray}
  c \sync d \SSeq \Magic &= &(c \sync d \SSeq \Magic) \SSeq \Abort \label{distribute-infeasible-suffix-abort}
\end{eqnarray}
by
\[
 \begin{array}{l}
  c \sync d \SSeq \Magic
 \Equals*[as $\Magic = \Magic \SSeq \Abort$ from (\refprop{seq-distr-right})]
  (c \SSeq \Nil) \sync (d \SSeq \Magic \SSeq \Abort)
 \Refsto*[by interchanging synchronisation and sequential (\refprop{sync-interchange-seq})]
  (c \sync d\SSeq \Magic) \SSeq (\Nil \sync \Abort)
 \Equals*[by the assumption that $\sync$ is abort-strict]
  (c \sync d\SSeq \Magic) \SSeq \Abort
 \Refsto*[as $\Abort \refsto \Nil$]
  c \sync d\SSeq \Magic
 \end{array}
 \]
from which we can conclude from (\refprop{seq-annihilation-left}) that for any $d_1,d_2\in \Commands$
\begin{eqnarray}
  (c \sync d \SSeq \Magic)\SSeq d_1
  =
  (c \sync d \SSeq \Magic)\SSeq \Abort \SSeq d_1
  =
  (c \sync d \SSeq \Magic)\SSeq \Abort \SSeq d_2
  =
  (c \sync d \SSeq \Magic)\SSeq d_2
  \label{infeasible-suffix-annihilates}
\end{eqnarray}
The result then follows in one direction from
\[
\begin{array}{l}
(c \sync d\SSeq \Magic)
\Equals*[$\Nil$ is the identity of sequential]
(c \SSeq \Nil) \sync (d \SSeq \Magic)
\Refsto*[by interchanging synchronisation and sequential (\refprop{sync-interchange-seq})]
(c \sync d) \SSeq (\Nil \sync \Magic)
\Equals*[using (\refprop{sync-nil-atomic})]
(c \sync d) \SSeq \Magic
\end{array}
\]
and then in the other from
\[
 \begin{array}{l}
  (c \sync d) \SSeq \Magic
\Refsto*[$\Nil \refsto \Magic$ and monotonicity]
  (c \sync d\SSeq \Magic) \SSeq \Magic
\Equals*[using (\ref{infeasible-suffix-annihilates})]
  (c \sync d\SSeq \Magic) \SSeq \Nil
\Equals*[$\Nil$ is sequential identity]
  (c \sync d\SSeq \Magic)
 \end{array}
\]
\qed
\end{proof}

\begin{lemmax}[conjoin-more-steps]
For $i,k \in\nat$, if $i \leq k$ then
\begin{eqnarray*}
(c \together \cstepd^{i}) \SSeq d \together \cstepd^{k}   & = & (c \together \cstepd^i) \SSeq (d \together \cstepd^{k-i}) \\
(c \together \cstepd^{i}) \SSeq d \together \Inf{\cstepd} & = & (c \together \cstepd^i) \SSeq (d \together \Inf{\cstepd})
\end{eqnarray*}
\end{lemmax}

\begin{proof}
Using assumptions $i\in \nat$ and $i \leq k$, we have that
$\cstepd^{k} = \cstepd^{i}\SSeq \cstepd^{k-i}$ and $\Inf{\cstepd} = \cstepd^{i}\SSeq \Inf{\cstepd}$,
and the results follows directly from (\refprop{conjoin-sync-initial}).
\qed
\end{proof}

\begin{lemmax}[conjoin-less-steps]
For $i,k\in \nat$, if $i > k$ then 
\begin{eqnarray*}
(c \together \cstepd^{i})   \SSeq d \together \cstepd^{k}  & = & (c \together \cstepd^k) \SSeq \Magic\\
(c \together \Inf{\cstepd}) \SSeq d \together \cstepd^{k} & = & (c \together \cstepd^k) \SSeq \Magic
\end{eqnarray*}
\end{lemmax}

\begin{proof}
So that we can consider both properties in the proof simultaneously
(i.e. for $i\in \nat$ and $i=\infty$) we define $\nat_{\infty} \sdefs
\nat \cup \{\infty\}$, and write $i-k$ as shorthand for $\infty$ when
$i=\infty$ and $k\in \nat$.
First we prove the simpler result that for any command $d$, $i\in
\nat_{\infty}$, if $i > k$ then
\begin{eqnarray}
\cstepd^{i}\SSeq d \together \cstepd^{k} & = & \cstepd^{k}\SSeq \Magic \label{conjoin-less-steps-simple}
\end{eqnarray}
because
\begin{displaymath}
\begin{array}{l}
\cstepd^{i}\SSeq d \together \cstepd^{k}
\Equals*[assumption $i>k$ letting $i{-}k=\infty$ if $i{=}\infty$, $\Nil$ is the identity of sequential]
\cstepd^{k}\SSeq (\cstepd^{i-k} \SSeq d) \together \cstepd^{k}\SSeq \Nil
\Equals*[by property (\refprop{sync-interchange-seq-atomic}) $k$ times]
(\cstepd \together \cstepd)^{k} \SSeq (\cstepd^{i-k} \SSeq d \together \Nil)
\Equals*[weak conjunction is idempotent (\refprop{conjoin-idempotent})]
\cstepd^{k} \SSeq (\cstepd^{i-k} \SSeq d \together \Nil)
\Equals*[from (\refprop{sync-nil-atomic}) because $i>k$]
\cstepd^{k} \SSeq \Magic
\end{array}
\end{displaymath}
By applying property (\ref{conjoin-less-steps-simple}) twice we have
that for any command $d$, $i\in \nat_{\infty}$, if $i > k$ then
\begin{eqnarray}
  \cstepd^{i}\SSeq d \together \cstepd^{k}
  = \cstepd^{k}\SSeq \Magic
  = \cstepd^{i}\SSeq \Nil \together \cstepd^{k}
  = \cstepd^{i} \together \cstepd^{k} \label{conjoin-less-steps-simple-ik}
\end{eqnarray}
We then reason that for $i\in \nat_{\infty}$
\begin{eqnarray}
(c \together \cstepd^{i}) \SSeq d & = & (c \together \cstepd^{i})\SSeq d \together (\cstepd^{i} \SSeq \Om{\cstepd}) \label{at-least-i-steps}
\end{eqnarray}
That is to say, the command $(c \together \cstepd^{i})\SSeq d$ takes
at least $i$ steps. For $i\in \nat$ this holds by property
(\refprop{conjoin-sync-initial}) and using the fact that weak
conjunction is idempotent (\refprop{conjoin-idempotent}), with
identity $\Chaos = \Om{\cstepd}$; and for $i=\infty$ we use
\Lemma{infinite-annihilates}, property $\Inf{c} = \Inf{c}\SSeq d$ for
any $c$ and $d$, and weak conjunction is idempotent.

Assuming $i\in \nat_{\infty}$ and $i>k$, and using these properties we then show that
\begin{displaymath}
\begin{array}{l}
(c \together \cstepd^{i}) \SSeq d \together \cstepd^{k}
\Equals*[using (\ref{at-least-i-steps})]
(c \together \cstepd^{i}) \SSeq d \together (\cstepd^{i} \SSeq \Om{\cstepd}) \together \cstepd^{k}
\Equals*[using the assumption $i>k$ and (\ref{conjoin-less-steps-simple-ik})]
(c \together \cstepd^{i}) \SSeq d \together (\cstepd^{k} \together \cstepd^{i})
\Equals*[using $\Nil$ is the identity of sequential composition and (\refprop{conjoin-sync-initial})]
(c \together \cstepd^{i} \together \cstepd^{k}) \SSeq (d \together \Nil)
\Equals*[using the assumption $i>k$ and (\ref{conjoin-less-steps-simple})]
(c \together \cstepd^{k}\SSeq \Magic) \SSeq (d \together \Nil)
\Equals*[using \Lemma{distribute-infeasible-suffix}, and $\Magic\SSeq c = \Magic$ from (\refprop{seq-distr-right})]
(c \together \cstepd^{k})\SSeq \Magic
\end{array}
\end{displaymath}
\qed
\end{proof}

\pagebreak[3]

\begin{lemmax}[split-finite-iter]
For $k\in\nat$, $(c \SSeq d) \together \cstepd^{k}  =
\Nondet_{i \leq k} ((c \together \cstepd^{i})\SSeq (d \together \cstepd^{k-i}))$~.
\end{lemmax}

\begin{proof}
\begin{displaymath}
\begin{array}{l}
c \SSeq d \together \cstepd^{k}
\Equals*[identity of weak conjunction is $\Om{\cstepd}$]
(c \together \Om{\cstepd}) \SSeq d \together \cstepd^{k}
\Equals*[decomposition of iteration from (\refprop{isolation}) and (\refprop{finite-iteration})]
(c \together (\Nondet_{i\in \nat_{\infty}}\cstepd^{i})) \SSeq d \together \cstepd^{k}
\Equals*[distributing choices]
\Nondet_{i\in \nat_{\infty}} ((c \together \cstepd^{i})\SSeq d \together \cstepd^{k})
\Equals*[using \Lemma{conjoin-more-steps} and \Lemma{conjoin-less-steps}]
\Nondet_{i \leq k} ((c \together \cstepd^{i})\SSeq (d \together \cstepd^{k-i}))
\nondet
(c \together \cstepd^{k})\SSeq \Magic
\Equals*[using $(c \together \cstepd^{k})\SSeq (d \together \cstepd^{0}) \refsto (c \together \cstepd^{k})\SSeq \Magic$ by monotonicity]
\Nondet_{i \leq k} ((c \together \cstepd^{i})\SSeq (d \together \cstepd^{k-i}))
\end{array}
\end{displaymath}
\qed
\end{proof}

The proof of the following lemma is similar to that for
\Lemma{split-finite-iter}.

\begin{lemmax}[split-infinite-iter]
If $c = c \together \Finite$, then
\begin{eqnarray*}
(c \SSeq d) \together \Inf{\cstepd}  &=&  \Nondet_{i\in \nat} (c \together \cstepd^i) \SSeq (d \together \Inf{\cstepd})~.
\end{eqnarray*}
\end{lemmax}

\begin{lemmax}[parallel-iter]
For $i\in \nat_{\infty}$, 
\begin{eqnarray*}
(c \together \cstepd^{i}) \parallel d  &=&  (c \parallel d) \together \cstepd^i \\
(c \together \cstepd^{\infty}) \parallel d  &=&  (c \parallel d) \together \cstepd^{\infty} \\
(c \together \Fin{\cstepd}) \parallel d &=& (c \parallel d) \together \Fin{\cstepd}
\end{eqnarray*}
  
\end{lemmax}

\begin{proof}
By (\refprop{conjoin-par-finite}) or (\refprop{conjoin-par-infinite})
and associativity and commutativity of parallel we have that
\begin{eqnarray}
(c \together \cstepd^i) \parallel d
=
(c \parallel \cestepd^i) \parallel d
=
(c \parallel d) \parallel \cestepd^i
=
(c \parallel d) \together \cstepd^i ~.
\label{parallel-iter-fixed}
\end{eqnarray} 
Using this to show the second property we then have
\begin{displaymath}
\begin{array}{l}
(c \together \Fin{\cstepd}) \parallel d
\Equals*[decomposition of finite iteration (\refprop{finite-iteration})]
(c \together (\Nondet_{i\in \nat} \cstepd^{i})) \parallel d
\Equals*[distributing choice]
\Nondet_{i\in \nat} ((c \together \cstepd^{i}) \parallel d)
\Equals*[using the first property (\ref{parallel-iter-fixed})]
\Nondet_{i\in \nat} ((c \parallel d) \together \cstepd^{i})
\Equals*[re-distributing the nondeterministic choice]
((c \parallel d) \together (\Nondet_{i\in \nat} \cstepd^{i})
\Equals*[re-composing finite iteration]
(c \parallel d) \together \Fin{\cstepd}
\end{array}
\end{displaymath}
\qed
\end{proof}

\begin{lemmax}[parallel-split]
For commands $c$ and $d$,
\begin{eqnarray*}
c \parallel d & = & \Nondet_{k\in \nat_{\infty}} ((c \together \cstepd^{k}) \parallel (d \together \cstepd^{k}))  
\end{eqnarray*}
and if either $c = c \together \Fin{\cstepd}$ or $d= d \together \Fin{\cstepd}$, then 
\begin{eqnarray*}
c \parallel d & = & \Nondet_{k\in \nat} ((c \together \cstepd^{k}) \parallel (d \together \cstepd^{k}))~.
\end{eqnarray*}
\end{lemmax}

\begin{proof}
We prove the first result, the proof for the second is similar. 
\begin{displaymath}
\begin{array}{l}
\Nondet_{k\in \nat_{\infty}} ((c \together \cstepd^{k}) \parallel (d \together \cstepd^{k}))
\Equals*[using \Lemma{parallel-iter} twice]
\Nondet_{k\in \nat_{\infty}} (c \parallel \cestepd^{k} \parallel d \parallel \cestepd^{k})
\Equals*[associativity and $\cestepd^{k} \parallel \cestepd^{k} = \cestepd^{k}$ by (\refprop{sync-interchange-seq-atomic}) and (\refprop{sync-env})]
\Nondet_{k\in \nat_{\infty}} ((c \parallel d) \parallel \cestepd^{k})
\Equals*[distributing choice]
(c \parallel d) \parallel \Nondet_{k\in \nat_{\infty}} \cestepd^{k}
\Equals*[decomposition of iteration from (\refprop{isolation}) and (\refprop{finite-iteration})]
(c \parallel d) \parallel \Skip
\Equals*[$\Skip$ is the identity of parallel]
c \parallel d
\end{array}
\end{displaymath}
\qed
\end{proof}

\begin{lemmax}[parallel-split-sequential]
If either $c_1= c_1 \together \Fin{\cstepd}$ or $d_1 = d_1 \together \Fin{\cstepd}$,
\begin{eqnarray*}
(c_1 \parallel d_1) \SSeq (c_2 \parallel d_2) & = &
\Nondet_{k\in \nat, k'\in \nat_{\infty}}
(c_1 \together \cstepd^{k})\SSeq (c_2 \together \cstepd^{k'})
\parallel
(d_1 \together \cstepd^{k})\SSeq (d_2 \together \cstepd^{k'})
\end{eqnarray*}
\end{lemmax}

\begin{proof}
\begin{displaymath}
\begin{array}{l}
(c_1 \parallel d_1) \SSeq (c_2 \parallel d_2)
\Equals*[using \Lemma{parallel-split} twice]
(\Nondet_{k\in \nat}((c_1 \together \cstepd^k) \parallel (d_1 \together \cstepd^k)))
\SSeq
(\Nondet_{k\in \nat_{\infty}} ((c_2 \together \cstepd^k) \parallel (d_2 \together \cstepd^k)))
\Equals*[distributing choices]
\Nondet_{k\in \nat, k'\in \nat_{\infty}}
((c_1 \together \cstepd^k)\parallel (d_1 \together \cstepd^k))
\SSeq
((c_2 \together \cstepd^{k'})\parallel (d_2 \together \cstepd^{k'}))
\Equals*[by (\refprop{sync-initial})]
\Nondet_{k\in \nat, k'\in \nat_{\infty}}
((c_1 \together \cstepd^k) \SSeq (c_2 \together \cstepd^{k'}))
\parallel
((d_1 \together \cstepd^k) \SSeq (d_2 \together \cstepd^{k'}))
\end{array}
\end{displaymath}
\qed
\end{proof}

\pagebreak[4]

\begin{lemmay}[sync-termination]
For commands $c$ and $d$ such that $c=c\together \Finite$ and $d = d\together \Finite$,
\begin{eqnarray*}
(c \SSeq \Fin{a} \parallel d \SSeq \Fin{b}) \SSeq (\Om{a} \parallel \Om{b})
& = &
c \SSeq \Om{a} \parallel d \SSeq \Om{b}
\end{eqnarray*}
\end{lemmay}

\begin{proof}
\begin{displaymath}
\begin{array}{l}
(c \SSeq \Fin{a} \parallel d \SSeq \Fin{b}) \SSeq (\Om{a} \parallel \Om{b})
\Equals*[using \Lemma{parallel-split-sequential}]
\Nondet_{k\in\nat, k'\in \nat_{\infty}}
((c \SSeq \Fin{a})\together \cstepd^k)\SSeq (\Om{a} \together \cstepd^{k'})
\parallel
((d \SSeq \Fin{b})\together \cstepd^k)\SSeq (\Om{b} \together \cstepd^{k'})
\Equals*[using \Lemma{split-finite-iter} and distributing choice]
\Nondet_{k\in\nat, k'\in \nat_{\infty}} \\
(\Nondet_{i\in\nat}^{i\leq k}(c \together \cstepd^{i}) \SSeq (\Fin{a} \together \cstepd^{k-i})
\SSeq (\Om{a} \together \cstepd^{k'}))
\parallel
(\Nondet_{j\in\nat}^{j\leq k}(d \together \cstepd^{j}) \SSeq (\Fin{b} \together \cstepd^{k-j})
\SSeq (\Om{b} \together \cstepd^{k'}))
\Equals*[simplify sequential composition of atomic steps]
\Nondet_{k\in\nat, k'\in \nat_{\infty}}
(\Nondet_{i\in\nat}^{i\leq k}(c \together \cstepd^{i}) \SSeq (\Om{a} \together \cstepd^{k+k'-i}))
\parallel
(\Nondet_{j\in\nat}^{j\leq k}(d \together \cstepd^{j}) \SSeq (\Om{b} \together \cstepd^{k+k'-j})
\Equals*[simplify]
\Nondet_{x\in\nat_{\infty}}
(\Nondet_{i\in\nat}^{i\leq x}(c \together \cstepd^{i}) \SSeq (\Om{a} \together \cstepd^{x-i}))
\parallel
(\Nondet_{j\in\nat}^{j\leq x}(d \together \cstepd^{j}) \SSeq (\Om{b} \together \cstepd^{x-j})
\Equals*[using \Lemma{split-finite-iter} and \Lemma{split-infinite-iter}]
\Nondet_{x\in \nat_{\infty}}
((c \SSeq \Om{a}) \together \cstepd^{x})
\parallel
((d \SSeq \Om{b}) \together \cstepd^{x})
\Equals*[using \Lemma{parallel-split}]
c  \SSeq \Om{a}
\parallel
d  \SSeq \Om{b}
\end{array}
\end{displaymath}
\qed
\end{proof}

\begin{lemmay}[fair-par-fair-expand]
$\Just \parallel \Just ~=~ \Fin{\cestepd} \SSeq (\Nil \nondet \cpstepd \SSeq (\Just \parallel \Just))$
\end{lemmay}

\begin{proof}
\[
 \begin{array}{l}
  \Fin{\cestepd} \SSeq \Om{(\cpstepd \SSeq \Fin{\cestepd})} \parallel
  \Fin{\cestepd} \SSeq \Om{(\cpstepd \SSeq \Fin{\cestepd})}
 \Equals*[by \Lemma{finite-finite-prefix} and (\refprop{sync-env})]
  \Fin{\cestepd} \SSeq
  ((\Om{(\cpstepd \SSeq \Fin{\cestepd})} \parallel \Om{(\cpstepd \SSeq \Fin{\cestepd})}) \nondet
   (\Om{(\cpstepd \SSeq \Fin{\cestepd})} \parallel \cestepd \SSeq \Fin{\cestepd} \SSeq \Om{(\cpstepd \SSeq \Fin{\cestepd})}))
 \Equals*[by \Lemma{iterate-pi-par-pi}; \Lemma{iterate-pi-sync-atomic} and (\refprop{sync-env})]
  \Fin{\cestepd} \SSeq 
  (\Nil \nondet  \SSeq \cpstepd (\Fin{\cestepd} \SSeq \Om{(\cpstepd \SSeq \Fin{\cestepd})} \parallel 
   \Fin{\cestepd} \SSeq \Om{(\cpstepd \SSeq \Fin{\cestepd})}))
  \Equals*[by the definition of $\Just$ (\refdef{fair})]
    \Fin{\cestepd} \SSeq (\Nil \nondet \cpstepd \SSeq (\Just \parallel \Just))
 \end{array}
\]
\qed
\end{proof}

\begin{lemmay}[fair-par-chaos-expand]
$\Just \parallel \Chaos = \Fin{\cestepd} \SSeq (\Nil \nondet \cpstepd \SSeq (\Just \parallel \Chaos))$
\end{lemmay}

\begin{proof}
\[
 \begin{array}{l}
  \Just \parallel \Chaos
 \Equals*[by definitions of $\Just$ (\refdef{fair}) and $\Chaos$ (\refdef{chaos}) and (\refprop{omega-decomposition})]
  \Fin{\cestepd} \SSeq \Om{(\cpstepd \SSeq \Fin{\cestepd})} \parallel \Om{\cestepd} \SSeq \Om{(\cpstepd \SSeq \Om{\cestepd})}
 \Equals*[by \Lemma{finite-omega-prefix} and (\refprop{sync-env})]
  \Fin{\cestepd} \SSeq 
    ((\Om{(\cpstepd \SSeq \Fin{\cestepd})} \parallel \Om{(\cpstepd \SSeq \Om{\cestepd})}) \nondet 
     (\Om{(\cpstepd \SSeq \Fin{\cestepd})} \parallel \cestepd \SSeq \Om{\cestepd} \SSeq \Om{(\cpstepd \SSeq \Om{\cestepd})}) \nondet 
     (\cestepd \SSeq \Fin{\cestepd} \SSeq \Om{(\cpstepd \SSeq \Fin{\cestepd})} \parallel \Om{(\cpstepd \SSeq \Om{\cestepd})}))
 \Equals*[by \Lemma{iterate-pi-par-pi} and \Lemma{iterate-pi-sync-atomic} and (\refprop{sync-env})]
  \Fin{\cestepd} \SSeq 
    (\Nil \nondet 
     \cpstepd \SSeq (\Fin{\cestepd} \SSeq \Om{(\cpstepd \SSeq \Fin{\cestepd})} \parallel \Om{\cestepd} \SSeq \Om{(\cpstepd \SSeq \Om{\cestepd})}) \nondet
     \cpstepd \SSeq (\Fin{\cestepd} \SSeq \Om{(\cpstepd \SSeq \Fin{\cestepd})} \parallel \Om{\cestepd} \SSeq \Om{(\cpstepd \SSeq \Om{\cestepd})}))
 \Equals*[by (\refprop{omega-decomposition}) and definitions (\refdef{fair}) and (\refdef{chaos})]
  \Fin{\cestepd} \SSeq (\Nil \nondet \cpstepd \SSeq (\Just \parallel \Chaos))
    \quad\qed
 \end{array}
\]
\end{proof}

\printindex
\end{document}